\documentclass[reprint,prd,a4paper,notitlepage]{revtex4-1}
\usepackage[pdftex]{graphicx}
\usepackage{inputenc}
\usepackage{amsmath}
\usepackage{amssymb}
\usepackage{subfigure}
\usepackage{multirow}

\usepackage{color}

\usepackage{hyperref}

\inputencoding{latin9}

\newcommand{\nc}{\newcommand}

\def\lsim{\; \raise0.3ex\hbox{$<$\kern-0.75em
      \raise-1.1ex\hbox{$\sim$}}\; }
\def\gsim{\; \raise0.3ex\hbox{$>$\kern-0.75em
      \raise-1.1\textmd{}ex\hbox{$\sim$}}\; }

\nc{\be}[1]{\begin{equation}\mbox{$\label{#1}$}}
\nc{\bea}[1]{\begin{eqnarray} \mbox{$\label{#1}$}}
\nc{\Section}[2]{\section{#2}\label{#1}}
\nc{\Bibitem}[1]{\bibitem{#1}}
\nc{\Label}[1]{\label{#1}}
\nc{\ie}{{\em i.e. }}
\nc{\eg}{{\em e.g. }}
\nc{\eea}{\end{eqnarray}}
\nc{\ee}{\end{equation}}
\nc{\w}{\omega}
\begin{document}

\title{Curvaton preheating revisited}

\author{J. Sainio}
\thanks{jani.sainio@utu.fi}
\affiliation{Department of Physics and Astronomy,
University of Turku, FIN-20014 Turku, FINLAND}
\date{\today}

\begin{abstract}


We study the thermalization process in the self-interacting curvaton preheating scenario. We solve the evolution of the system with classical lattice simulations with a recently released symplectic PyCOOL program during the resonance and the early thermalization periods and compare the results to the inflaton preheating. After this we calculate the generated non-gaussianity with the $\Delta N$ formalism and the separate universe approximation by running a large number of simulations with slightly different initial values. The results indicate a high level of non-gaussianity. We also use this paper to showcase the various post-processing functions included with the PyCOOL program that is available from \url{https://github.com/jtksai/PyCOOL}.

\end{abstract}

\maketitle

\section{Introduction}


The curvaton mechanism \cite{Linde:1996gt,Enqvist:2001zp,Lyth:2001nq,Bartolo:2002vf,Moroi:2002rd,Dimopoulos:2003ss,Moroi:2001ct} is a much studied alternative to the standard inflationary paradigm for the origin of the observed primordial perturbations. The curvaton field is assumed to be light and subdominant during the inflation process and its contribution to the energy density is significant only moments before its decay. This allows the inflation potential to have more natural properties \cite{Dimopoulos:2002kt} compared to the single field scenario while still leading to adiabatic perturbations consistent with the current observational data \cite{Spergel:2006hy,Komatsu:2008hk}.

Reheating of the universe is an important part of the early universe cosmology (for a review cf. \cite{Allahverdi:2010xz}). In the curvaton scenario it is most often assumed that the curvaton field decays perturbatively into lighter degrees of freedom once the Hubble parameter is of the order of curvaton decay width $\Gamma$ and thermalizes with the radiation that originates from the inflaton. It is however also possible that the universe reheated through a rapid and rather violent preheating process. This parametric resonance was studied in ref. \cite{Enqvist:2008be} in the curvaton scenario and the main conclusion was that in general it is quite similar to the preheating of the inflaton field. In ref. \cite{Chambers:2009ki} it was further found that the curvaton resonance can lead to very high levels non-gaussianity.

The curvaton potential in most of these studies is assumed to be of a quadratic type. As was noted in refs. \cite{Enqvist:2009zf,Kawasaki:2011pd} any deviations from this shape can lead to significant differences in the end results, especially in the level of generated non-gaussianity. Whereas these studies were limited to the perturbative regime in the present paper we expand this analysis to the non-linear preheating process. We limit the potential function of the curvaton to the typical quadratic type with additional quartic self-interactions. We also assume that the curvaton field does not couple to other scalar fields in contrast to refs. \cite{Enqvist:2008be,Chambers:2009ki}.

We will study this self-interacting curvaton scenario with classical fields and lattice simulations from two different perspectives. We will first concentrate on the thermalization of the curvaton field during the resonance process. We will compare the results to the preheating of inflaton that has been studied thoroughly in \cite{{Kofman:1994rk,Kofman:1997yn,Prokopec:1996rr,Khlebnikov:1996mc,Khlebnikov:1996zt,Felder:2000hr,Frolov:2008hy,
Podolsky:2005bw,Felder:2006cc}} with analytical and numerical methods. After this we will concentrate on the calculation of generated non-gaussianity with the $\Delta N$ formalism \cite{Wands:2000dp}. This mainly numerical study will be done with the recently published symplectic PyCOOL program \cite{Sainio:2012mw} (available from \url{https://github.com/jtksai/PyCOOL}). We also use this paper to showcase the numerous post-processing functions included with the program.

This paper is organized as follows. In section II we present the curvaton model and the equations of motion. In section III we present the thermalization and non-gaussianity calculations and results. We conclude with a discussion in section IV.

\section{Curvaton model}


We model the curvaton field with a simple polynomial potential function with quartic self-interactions
\begin{equation} \label{curv-pot}
V_{\sigma} = \frac{1}{2}m_{\sigma}^2 \sigma^2 + \frac{1}{4}\lambda_{\sigma} \sigma^4
\end{equation}
where $\sigma$ is the curvaton field and $\lambda_{\sigma}$ is the coupling constant of the curvaton self-interactions.
Following \cite{Chambers:2009ki} we will set the initial energy density of the homogeneous radiation component equal to the potential energy of the inflaton
\begin{equation} \label{V phi}
V_{\phi} = \frac{1}{4}\lambda_{\phi} \phi^4
\end{equation}
where the coupling constant $\lambda_{\phi}$ is a free parameter and we set $\phi \sim m_{\textrm{Pl}}$, $m_{\textrm{Pl}}$ being the reduced Planck mass.
The curvaton field is effectively massless during inflation and hence it is require that
\begin{equation}
m_{\sigma}^2 + 3 \lambda_{\sigma} \sigma^2 \ll H_{*}^2,
\end{equation}
where $H_{*}$ is the value of the Hubble parameter during inflation. 

After the inflation ends the curvaton field stays almost constant until it starts to oscillate around its minimum when the Hubble parameter has decreased close to the value of the effective mass of curvaton. In the usual perturbative analysis the field would then start to decay into lighter particles once the Hubble parameter is roughly equal to the decay width of the curvaton.
In this paper we are however more interested in the non-perturbative analysis meaning that the interaction terms in the potential function (\ref{curv-pot}) now lead to the production of curvaton particles \cite{Greene:1997fu}. The curvaton field is assumed to decay perturbatively only long after the resonance period is over.

The closely related reheating process of a self-interacting inflaton field has been studied previously in refs. \cite{Kofman:1994rk,Greene:1997fu,Felder:2000hr,Khlebnikov:1996mc,Micha:2002ey} of which the last two use a similar interaction picture to this study.
We will assume that the quartic term dominates the curvaton potential and hence initially we set $\sigma > m_{\sigma}/\sqrt{\lambda_{\sigma}}$.
In the opposite case the reheating process does not happen and the curvaton field does not thermalize.

The creation of particles during this preheating has been studied extensively in \cite{Greene:1997fu} in the case of massless inflation and we will cite the most relevant results here. The mode equation of the curvaton particles with wave number $k$ can be written in terms of a more general Lam\'{e} equation
\begin{equation} \label{lame}
\tilde{\sigma}_k''+ \bigg(\kappa^2 +  \frac{g^2}{\lambda_{\sigma}} \;\textrm{cn}^2(\tilde{\eta}, \frac{1}{\sqrt{2}})\bigg)\tilde{\sigma}_k = 0,
\end{equation}
which is valid for $V_{\textrm{int}} = \frac{g^2}{2} \sigma^ 2 \chi^2$ type interaction terms where $\chi$ is another scalar field. This equation however reduces to the mode equation of the quartic self-interaction when $\frac{g^2}{\lambda_{\sigma}} = 3$ \cite{Greene:1997fu}. We have here also defined $\tilde{\sigma} = a \sigma$, used prime to denote time derivative with respect to the scaled conformal time which is defined in terms of the physical time $dt$ as $d \tilde{\eta} = a^{-1}\sqrt{\lambda_{\sigma}} \tilde{\sigma}_0 dt$ and $\textrm{cn}(\tilde{\eta}, \frac{1}{\sqrt{2}})$ is the Jacobi cosine function. We have also used a rescaled wave number $\kappa^2 = k^2/(\lambda_{\sigma}\tilde{\sigma}_0)$, where the rescaled curvaton amplitude $\tilde{\sigma}_0$ is measured at the end of inflation.
The values of $\kappa^2$ and $\frac{g^2}{\lambda_{\sigma}}$ that will lead to production of particles can be read from the corresponding stability/instability chart that can be found for example in \cite{Greene:1997fu}.

We will now approximate the mode equation of the massive self-interacting curvaton particles with equation (\ref{lame}) with $\frac{g^2}{\lambda_{\sigma}} = 3$ and we will also neglect the mass term which we assume to be small compared to the interaction term at least during the early part of the evolution. It is now easy to see from the stability/instability chart that the curvaton particles are produced at a band close to a rescaled momentum value of $\kappa^2 \simeq 1.6$ which in terms of the comoving momentum reads
\begin{equation} \label{eq:part-band}
k^2_p \simeq 1.6 \lambda_{\sigma}\tilde{\sigma}_0^2.
\end{equation}
This is the only momentum band and the other suitable momentum values correspond to single points \cite{Greene:1997fu}.

This resonant phase of particle production is followed by \cite{Khlebnikov:1996mc} a period of rescattering of the coherent curvaton mode ($k = 0$) and the created particles leading to a formation of multiple peaks in the spectrum of the field close to the harmonic frequencies of $k_p$. After this the system enters a regime of turbulent dynamics \cite{Allahverdi:2010xz} which is followed by a long period during which the field reaches the thermal state.

\subsection{Equations of motion}

We will solve the evolution of the system with a symplectic algorithm that is by design meant to conserve the energy of the system. Instead of solving the Euler-Lagrange equations of motion we will instead use the Hamiltonian equations that are split into explicitly integrable pieces. Note that prime in the following equations means derivative with respect to the conformal time $d \eta = a^{-1} dt$.

Starting from the Einstein-Hilbert action and after some simple Legendre transformations the Hamiltonian function of the system can be derived. Since we will solve the equations of motion numerically in a periodic comoving lattice the system needs to be discretized. We will use a second order accurate and fourth order isotropic stencils for the Laplacian operators derived in \cite{Patra:2005}. The discretized Hamiltonian function in conformal time then reads \cite{Sainio:2012mw}
\begin{equation} \label{eq:H}
\begin{aligned}
\mathcal{H} = & -\frac{ p_a^2 }{12 V_L m_{Pl}^2} + a^4\left(\frac{ V_L \rho_{\gamma,0}}{a^4} + \frac{ V_L \rho_{m,0}}{a^3}\right) \\
& + \sum_{i,\vec{x}} a^4 \Bigg( \frac{\pi_{i,\vec{x}}^2}{2 a^6} - \frac{\phi_{i,\vec{x}}D[\phi_{i,\vec{x}}](\vec{x})}{ 2 a^2 dx^2}\\
& \qquad \qquad \quad + V(\phi_{1,\vec{x}},...,\phi_{N,\vec{x}})\Bigg), \\
\end{aligned}
\end{equation}
where $p_a$ is the canonical momentum of the scale factor $a$, $V_L = n^3$ equals the size of the cubic lattice, $dx$ the spacing of the lattice $m_{Pl}$ is the reduced Planck mass, $\pi_{i,\vec{x}}$ is the canonical momentum of field $\phi_{i,\vec{x}}$ at position $\vec{x} = (x_1,x_2,x_3)$ in the lattice and $D[\phi_{i,\vec{x}}](\vec{x})$ is the Laplacian of field $\phi_{i,\vec{x}}$ at position $\vec{x}$. Note also that the summation is carried over all of the fields and all positions in the lattice. We have also incorporated homogeneous radiation $\rho_{\gamma,0}$ and non-relativistic matter $\rho_{m,0}$ components into the system. It can be easily seen \cite{Sainio:2012mw} that right hand side of the Hamiltonian (\ref{eq:H}) corresponds to the first Friedmann equation and is therefore conserved by the symplectic integrator.

The Hamiltonian equations related to this Hamiltonian now read for the scale parameter and its canonical momentum
\begin{equation} \label{eq:a}
\begin{aligned}
a' = \frac{\partial \mathcal{H}}{\partial p_a} = & - \frac{p_a}{ 6 V_L m_{Pl}^2}\\
p_a' = - \frac{\partial \mathcal{H}}{\partial a} = & \sum_{i,\vec{x}} a^3 \Bigg( \frac{\pi_{i,\vec{x}}^2}{a^6} + \frac{\phi_{i,\vec{x}}D[\phi_{i,\vec{x}}](\vec{x})}{ a^2 dx^2}\\
& - 4 V(\phi_{1,\vec{x}},...,\phi_{N,\vec{x}})\Bigg) - V_L \rho_{m,0}.
\end{aligned}
\end{equation}
Similarly the equations of motion of scalar field $i$ at grid point $\vec{z}$ read
\begin{equation} \label{eq:phi}
\begin{aligned}
\phi_{i,\vec{z}}' &= \frac{\partial \mathcal{H}}{\partial (\pi_{i,\vec{z}})} = \frac{\pi_{i,\vec{z}}}{a^2} \\
\pi_{i,\vec{z}}' &= -\frac{\partial \mathcal{H}}{\partial (\phi_{i,\vec{z}})} =  a^2 \frac{D[\phi_{i,\vec{z}}](\vec{z})}{ dx^2} - a^4 \frac{\partial V }{\partial (\phi_{i,\vec{z}})}
\end{aligned}
\end{equation}
which follow from equation (\ref{eq:H}) by differentiating under the summation sign and by summing over the coefficients $c_{d(\alpha)}$ of the discretized Laplacian. When integrating these equations we will first split them into explicitly integrable pieces and then use a suitable symplectic integrator.


\section{Numerical results}

\subsection{Initial values}

We use units where the reduced Planck mass $m_{\textrm{PL}}$ is set to one. We will also use a general mass $m = 10^{-9} m_{Pl}$ to define the lattice, the initial radiation energy density and the time step $d \eta$.
The physical time is measured in units of $m^{-1}$. We will use a conformal time step $d \eta = 0.001/m$ in the simulations and solve the evolution until $t_\textrm{phys} \, m \simeq 5000$.

The size of the lattice is limited by requirement that $L < 1/(aH)$ \ie the comoving horizon is larger than the comoving lattice at all times. Otherwise the assumption that has been used when deriving equation (\ref{eq:H}) that the metric is of the Friedmann-Robertson-Walker form $-ds^2 = a(\eta)^2(-d\eta^2 + d\vec{x}^2)$ would have to be adjusted to include also metric perturbations.

We used two different lattice sizes to run the simulations: the thermalization study was done with $256^3$ points whereas the non-gaussianity simulations were run on smaller $64^3$ lattices that are roughly 46 times faster to solve.
We set the comoving edge of the lattice to be $5/(3 m)$ in the thermalization simulations meaning that the comoving momenta are in the range $3.8 \, m < k < 380 \, m$ which we calculate with the effective wave number $k_\textrm{eff}$ instead of the magnitude of the wave vector. In the non-gaussianity calculations with a smaller lattice size we are compelled to reduce either the infrared or the ultraviolet resolution of the simulation. Simple numerical test runs show that the ultraviolet modes are more important for the evolution of the system to be consistent in these two cases. We have therefore used a comoving edge length $5/(12 m)$ meaning that the comoving momenta are in the range $15.1 \, m < k < 380 \, m$ in the non-gaussianity results.

The initial values for the curvaton field were chosen based on two criteria. In order for the quartic term to dominate in the potential function we simply set
\begin{equation}
\sigma_0 > \frac{m_{\sigma}}{\sqrt{\lambda_{\sigma}}}.
\end{equation}
We also want the momentum band where the particle creation happens, \ie Eq. (\ref{eq:part-band}), to be inside the lattice meaning that the parameters should be chosen such that
\begin{equation}
k_p \sim \sqrt{1.6 \lambda_{\sigma}}\sigma_0
\end{equation}
is neither too large nor too small.

With these criteria in mind we used the following values for the parameters: the mass of the curvaton is set to $1 \times 10^{-9}$, initial curvaton field value $\sigma_0 = 2 \times 10^{-4}$, curvaton self-interaction strength $\lambda_{\sigma} = 1 \times 10^{-7}$, initial radiation density $\lambda_{\phi} = 1 \times 10^{-16}$. The initial fractional energy density of the curvaton, $\Omega_{\sigma,0}$, corresponding to these values is of order $\sim 10^{-6}$ (see Figure \ref{fig8}). The momentum band where the particle creation happens is approximately at $k_p/m \sim 80$.

\subsection{Output variables}

Previous studies of the thermalization process after preheating have used a number of different variables to study and to illustrate the different phases of this process. The comoving number density and the related number density spectra are certainly some of the most interesting ones to use. There have been however a number of different definitions and ways to calculate these variables leading to slightly different results while the overall picture of the thermalization process stays the same.
In this study we use a definition for the number density $n_k$ that was previously used in LATTICEEASY \cite{Felder:2000hq}.
This is done by using conformal field values $\tilde{F}_{k,c} = a \tilde{F}_{k}$ and conformal time to write the equations of motion of the Fourier modes of the fields in the form of a simple harmonic oscillator
\begin{equation}
\tilde{F}_{k,c}'' + \tilde{\omega}_k^2 \tilde{F}_{k,c} = 0,
\end{equation}
where $\tilde{F}_{k} = L^{-3/2} F_k$ is the scaled Fourier mode of conformal field $a f$, $L$ is the comoving length of the lattice and
\begin{equation} \label{disp}
\tilde{\omega}_k^2 = k_\textrm{eff}^2 + a^2 m_\textrm{eff}^2 = k_\textrm{eff}^2 + a^2 \bigg\langle \frac{\partial^2 V}{\partial f^2} \bigg\rangle - \frac{a''}{a}
\end{equation}
is the comoving dispersion relation. Note that we have used the effective wave number $k_\textrm{eff}$ which is calculated from the discrete Fourier transform of the discretized Laplacian operator. The wave number is often however calculated with the magnitude of the wave vector $k^2 = k^2_\textrm{x} + k^2_\textrm{y} + k^2_\textrm{z}$ as is done for example in LATTICEEASY. This method might however lead to inaccurate number density results \cite{Stamatopoulos:2012fu} whereas the effective wave number takes properly the used discretization into account. We have also defined the effective mass $m_\textrm{eff}$ in equation (\ref{disp}) where the brackets denote an average over the lattice.
The number density of the scalar particles can be then written in terms of the scaled modes $\tilde{F}_{k}$ as
\begin{equation} \label{n_k}
n_k \equiv \frac{1}{2}\bigg(\tilde{\omega}_k |\tilde{F}_{k,c}|^2 + \frac{1}{\tilde{\omega}_k}|\tilde{F}'_{k,c}|^2\bigg),
\end{equation}
which is calculated by binning the data into spherical shells in the momentum space that are then averaged.
We will also study the time evolution of the number of particles in the comoving lattice
\begin{equation}
N(t) = \frac{1}{(2 \pi )^3}\int n_{k} d^3 k,
\end{equation}
which is calculated by summing over the non-averaged momentum bins.

We are also interested in various energy density related variables.
We first define the energy density spectra based on the number density equation (\ref{n_k}) as
\begin{equation} \label{rho_k}
\rho_k \equiv \omega_k n_k
\end{equation}
where now $\omega_k = \tilde{\omega}_k/a$ is the physical dispersion relation.
The energy density of a quanta at momentum $k$ then simply reads $\epsilon_k = k^2 \rho_k$ \cite{Podolsky:2005bw}.
The energy and the pressure density of a scalar field in position space are defined as
\begin{equation} \label{eq:rho-pres}
\begin{aligned}
\rho_i & \equiv  \frac{(\phi'_{i})^2}{2 a^2} + \frac{1}{2a^2}(\nabla\phi_{i})^2 + V(\phi_{i}) \\
P_i & \equiv \frac{(\phi'_{i})^2}{2 a^2} - \frac{1}{6a^2}(\nabla\phi_{i})^2 - V(\phi_{i}). \\
\end{aligned}
\end{equation}
We calculate the fractional energy densities from these expressions with
\begin{equation}
\Omega_i = \frac{\rho_i}{\rho_{\textrm{tot}}},
\end{equation}
where $\rho_{\textrm{tot}}$ now includes all of the scalar fields and the homogeneous radiation component.
The equation of state is derived from (\ref{eq:rho-pres})
\begin{equation} \label{eqs}
\omega_i = \frac{\langle P_i \rangle}{\langle \rho_i \rangle},
\end{equation}
where the brackets denote averaging over the lattice.

We are also interested in the statistical properties of the fields during the resonance process. In this study we will use the excess kurtosis which is defined as
\begin{equation} \label{kurtosis}
\gamma_2 = \frac{\mu_4}{\sigma^4} - 3,
\end{equation}
where $\mu_4$ is the fourth moment about the mean and $\sigma$ is the standard deviation (not to be confused with the curvaton field). This quantity is mainly used to gauge how much the distribution of the curvaton field deviates from a gaussian one for which it is identically zero. A large value of kurtosis generally indicates that the distribution has more mass in the tails.

\subsection{Thermalization results}

We use a fourth order symplectic integrator to solve the evolution of the system in conformal time. The output is calculated after a constant number of integration steps. The moving averages presented in the figures are calculated over these points meaning that when presented in physical time the length of the averaged period increases with time. We therefore use the term conformal moving average in the figures.

\begin{figure}[h]
{\includegraphics*[width=\columnwidth]{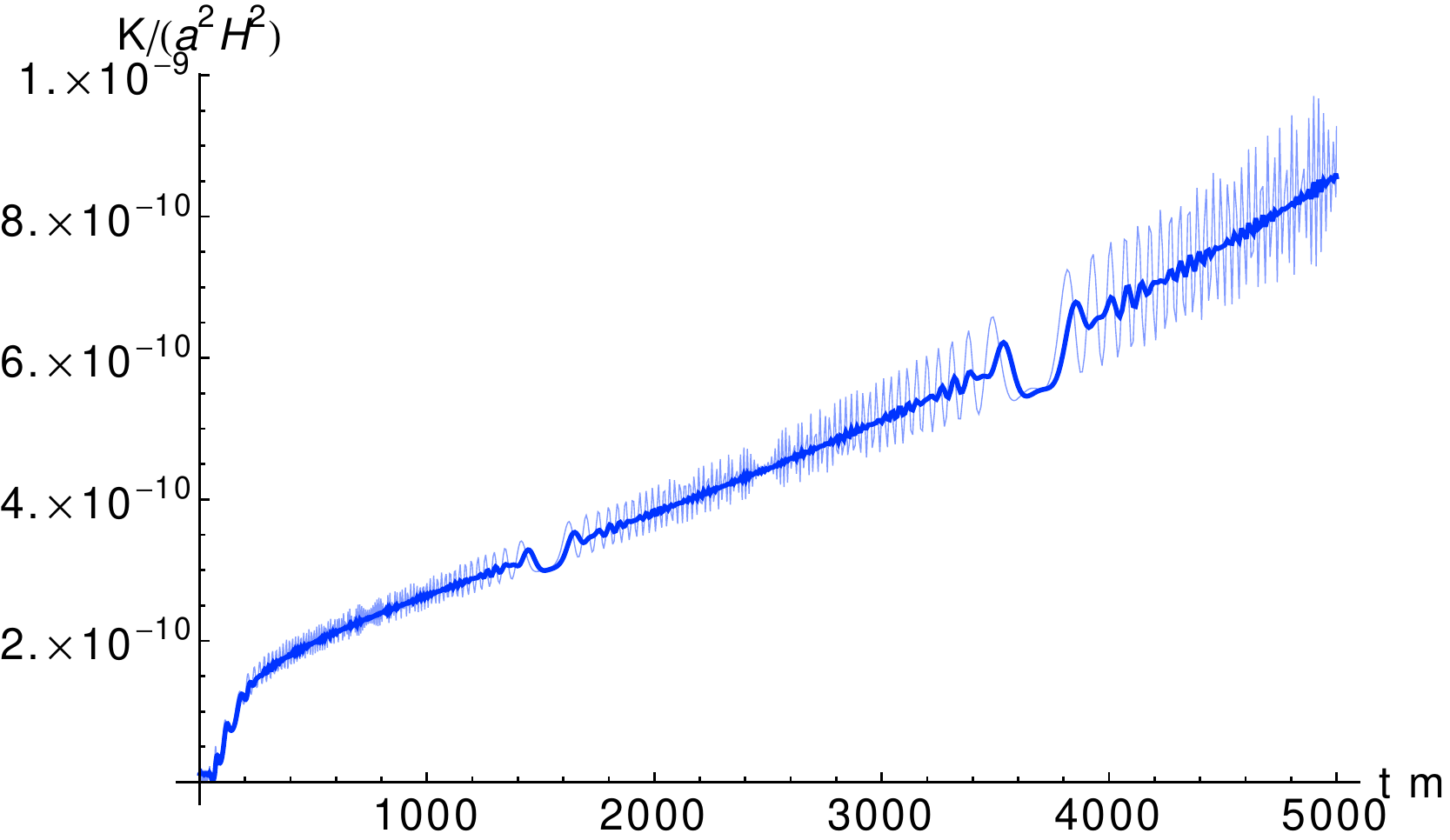}}
\caption{The evolution of the numerical error during the simulation (thin line) and its conformal moving average (thick line). Notice that the averages are calculated over the values at different output points which are written after a constant number of integration steps in conformal time. Therefore the physical time over which the moving average is calculated varies. Notice also that the large gaps at $t_\textrm{phys}/m \sim 1400$ and $t_\textrm{phys}/m \sim 3800$ in the graph are an artifact of this used sampling. Please see the online version of this article for color figures.}
\label{fig1}
\end{figure}

The numerical accuracy during the simulation is shown in Figure \ref{fig1} where we plot the absolute value of the residual curvature
\begin{equation} \label{eq:res-curv}
\frac{K}{a^2 H^2} = \Big|\frac{8\pi G \langle \rho \rangle }{3 H^2} - 1\Big|
\end{equation}
which we use to measure the conservation of Hamiltonian(\ref{eq:H}). As can be seen from the figure the algorithm is accurate to $10^{-10}$ level during the preheating phase. The error does increase with time but not substantially.

\begin{figure}[h]
{\includegraphics*[width=\columnwidth]{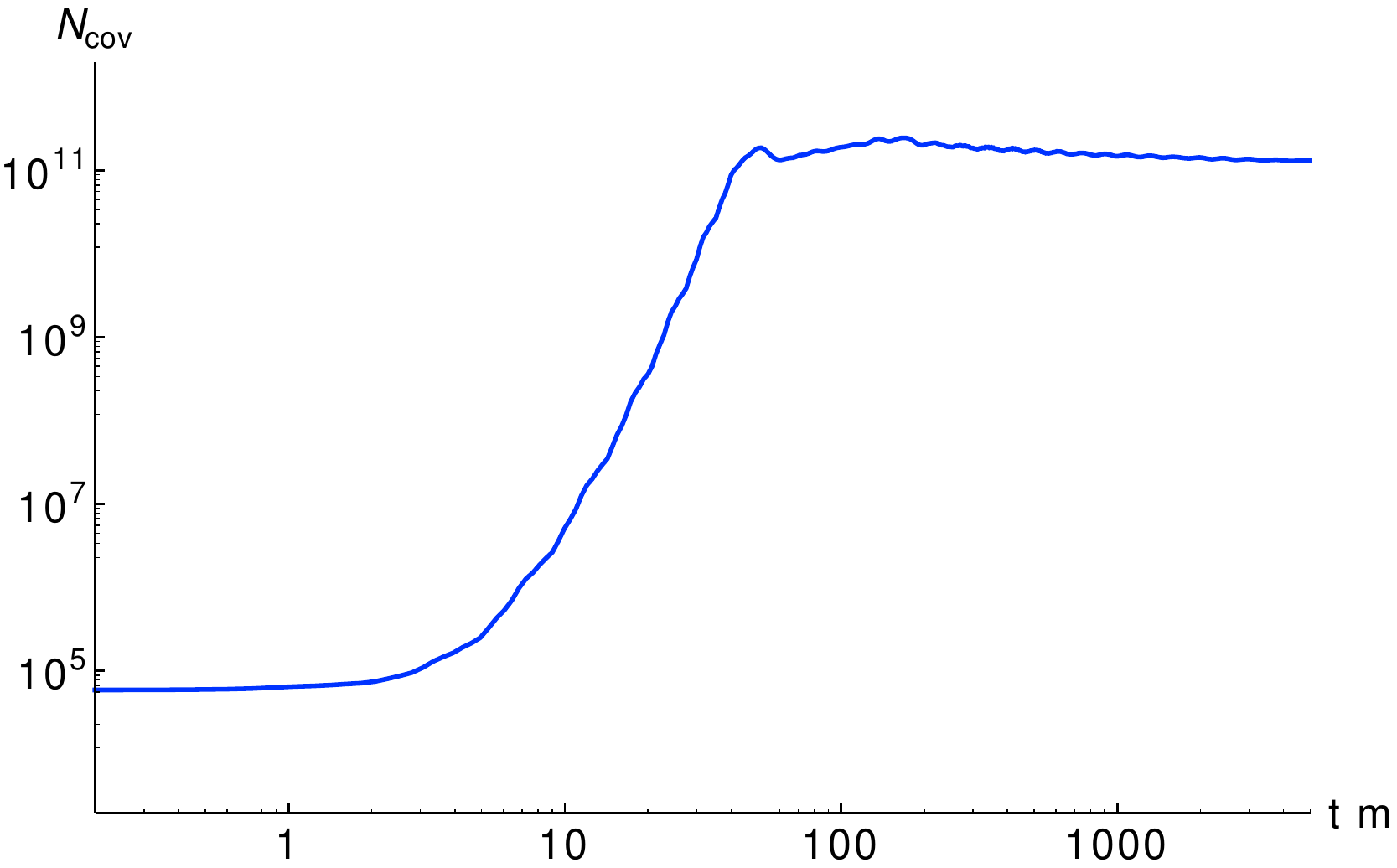}}
\caption{ The evolution of the comoving number density of the curvaton particles. Different phases of the process are clearly distinguishable: exponential resonance period at $1 \lesssim t_\textrm{phys}/m \lesssim 40$, the rescattering phase $40 \lesssim t_\textrm{phys}/m \lesssim 200$ and the final turbulence period.}
\label{fig2}
\end{figure}

The progress of the thermalization process is presented in Figure \ref{fig2} where we plot the comoving number density as a function of time. As can be seen from the figure the number density initially stays close to a constant but as the resonance process starts the number density begins to increase exponentially. At $t_\textrm{phys} \, m \sim 40$ the resonance ends and the system then enters the rescattering period. During this the number density reaches a short plateau phase after which it starts to gradually decrease mainly due to a lack of infrared resolution of the lattice. Overall the evolution of the number density is quite similar to the one witnessed in the chaotic inflation case \cite{Felder:2006cc}.

\begin{figure}[h]
{\includegraphics*[width=\columnwidth]{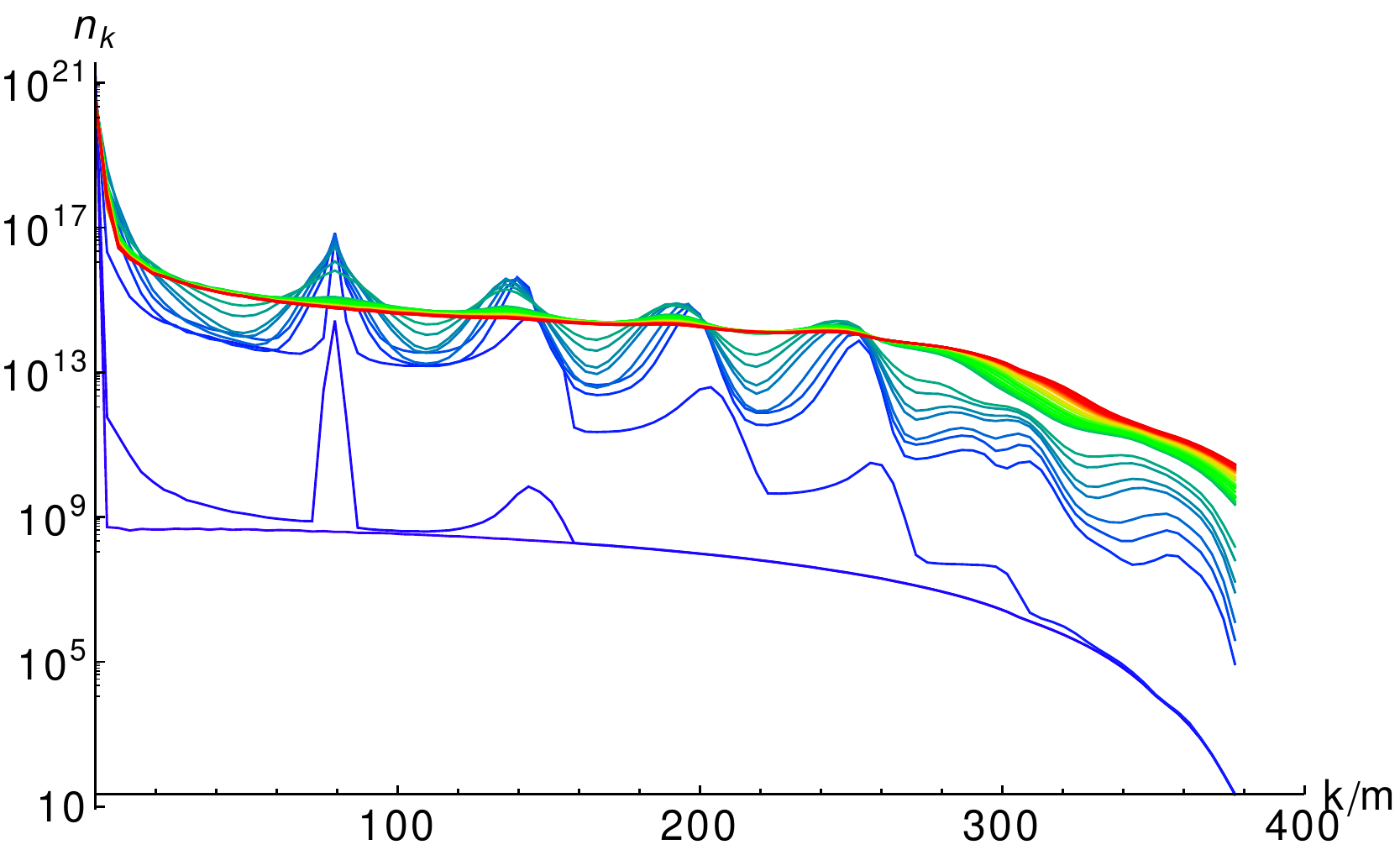}}
\caption{The evolution of the number density spectrum. Note that in the figure the color evolves with time and the red curves are calculated close to the end of the simulation whereas the blue ones (at the bottom) are evaluated at $t_\textrm{phys} \, m = 0$. The time difference between the spectra is roughly $t_\textrm{phys} \, m \simeq 20$. We have also included a power law fit $n_k \sim k^{-p}$ of the final spectrum with $p \simeq 3/2$ as a black dashed curve in the figure. Please see the online version of this article for color figures.}
\label{fig3}
\end{figure}

\begin{figure}[h]
{\includegraphics*[width=\columnwidth]{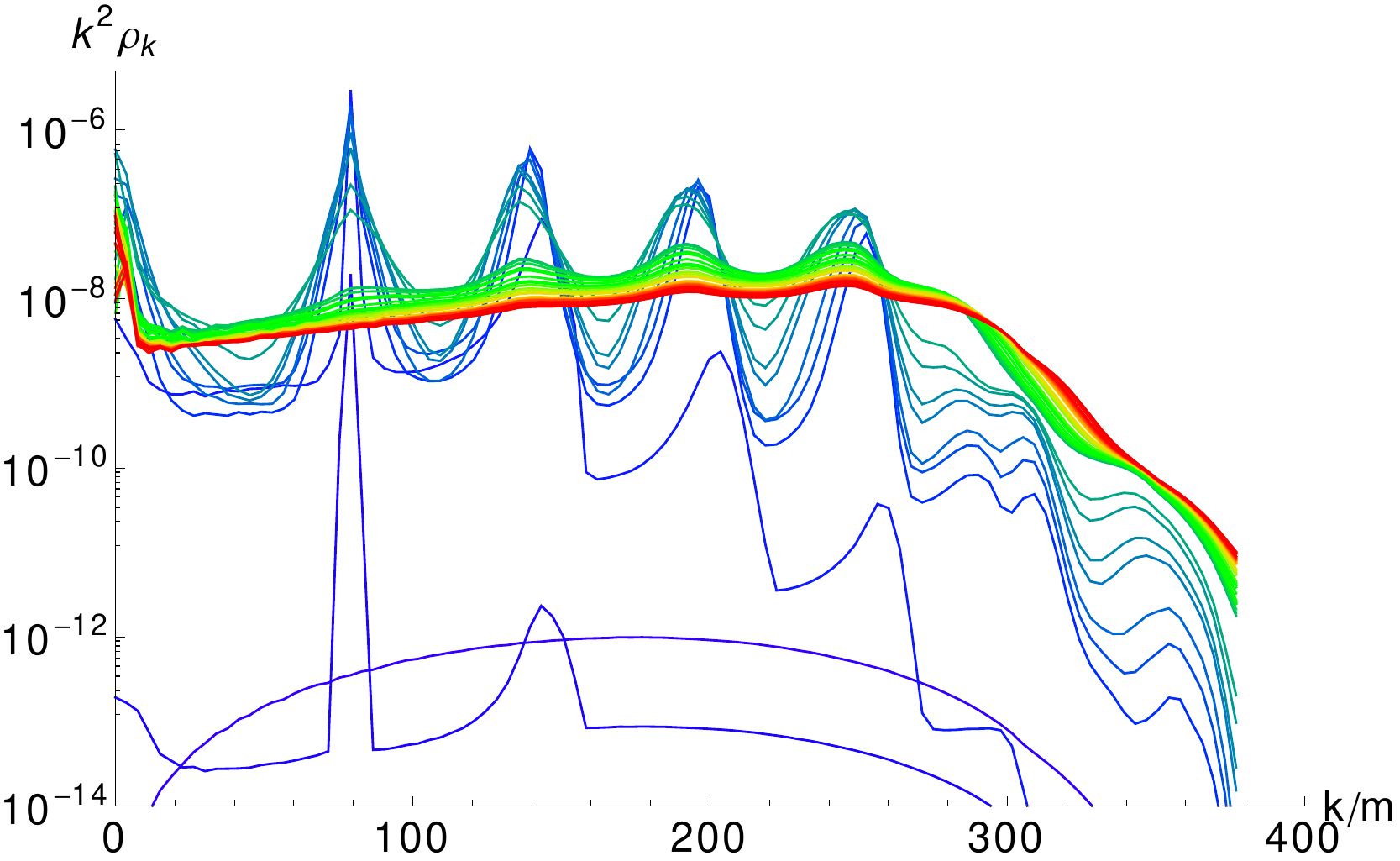}}
\caption{The evolution of the energy density spectrum of the quanta at momentum $k$. Note that in the figure the color evolves with time and the red curves are calculated close to the end of the simulation whereas the blue ones (at the bottom) are evaluated at $t_\textrm{phys} \, m = 0$. The time difference between the spectra is roughly $t_\textrm{phys} \, m \simeq 20$. Please see the online version of this article for color figures.
}
\label{fig4}
\end{figure}

Close inspection of the evolution of the number and energy density spectra however tells a very different story when compared to the chaotic inflation. In the broad parametric resonance of the chaotic inflation the preheating process is most efficient at creating particles with momentum values below a threshold value $k_{*}$ \cite{Felder:2000hr}. In terms of the energy density of the quanta at momentum $k$ the chaotic inflation potential usually leads to a spectrum with one peak at the inflaton particle energy spectrum that broadens with time and shifts to higher comoving momentum values with time \cite{Podolsky:2005bw}. In the self-interacting curvaton case the particle creation happens initially at the resonance band calculated in eq. (\ref{eq:part-band}) which can be seen in Figure \ref{fig3} as a formation of a clear peak at $k/m \sim 80$. This phase is however followed shortly by excitation of curvaton particles at a series of different bands indicating that the system has entered the rescattering period \cite{Khlebnikov:1996mc}.
Note that this part of the process is quite sensitive to the initial values: at larger initial radiation densities or smaller curvaton self-interaction values it is possible to stop this process before the other peaks start to form.

The shape of the number density spectrum at the end of the simulation is visible in Figure \ref{fig3} as a red curve. The observed peaks have leveled out except for small residual hills. Other notable feature is that the spectrum is elevated at smaller momentum values. This final shape also appears to be quite stable in the sense that it does not change considerably during the last stages of the simulation.
To compare this to a thermal boson spectrum 
we have fitted the data to the usual Rayleigh-Jeans approximation of the number density spectrum
\begin{equation}
n_k \approx \frac{T}{\omega_k-\mu}
\end{equation}
where $T$ is the comoving temperature of the boson field in thermal equilibrium and $\mu$ is the corresponding chemical potential. 
The best least squares fit (not shown in the figure) results in $\mu \simeq a \; m_\textrm{eff}$ and $T \gg a \;  m_{\textrm{Pl}}$ which strongly indicates that the system is non-thermal. A power law function $n_k \propto k^{-p}$ with $p \simeq 3/2$ seems to follow the shape of the spectrum more closely until an exponential cut-off at high momentum values. Similar result was previously presented in the case of self-interacting massless inflaton field in ref. \cite{Micha:2002ey} where the evolution of the spectra during the turbulence period was in addition found to be self-similar. Although we were unable to verify this with the curvaton model the results indicate that the curvaton is not at thermal equilibrium at the end of the simulation. 
Assuming that the eventual thermalization of the curvaton happens through the quartic interactions and that it is not coupled to other fields the corresponding decay rate reads
\begin{equation}
\Gamma \sim \frac{\lambda^2_{\sigma} \; m_{\sigma}}{4 \pi} \simeq 2 \times 10^{-6} \textrm{ GeV}.
\end{equation}
which leads to a rather low reheating temperature of a few MeV.

Another perspective to the resonance process can be seen in Figure \ref{fig4} where we plot the evolution of the energy density of the quanta at momentum $k$ with the quantity $k^2 \rho_k$. As is evident from the graph most of the curvaton particles are created at five different harmonic momentum bands. As time evolves the series of peaks smoothen as the thermalization process progresses and the energy density of the particles propagates toward higher momentum values. The final state in this case is very different from the one seen in the chaotic inflation \cite{Podolsky:2005bw}.

\begin{figure}[h]
{\includegraphics*[width=\columnwidth]{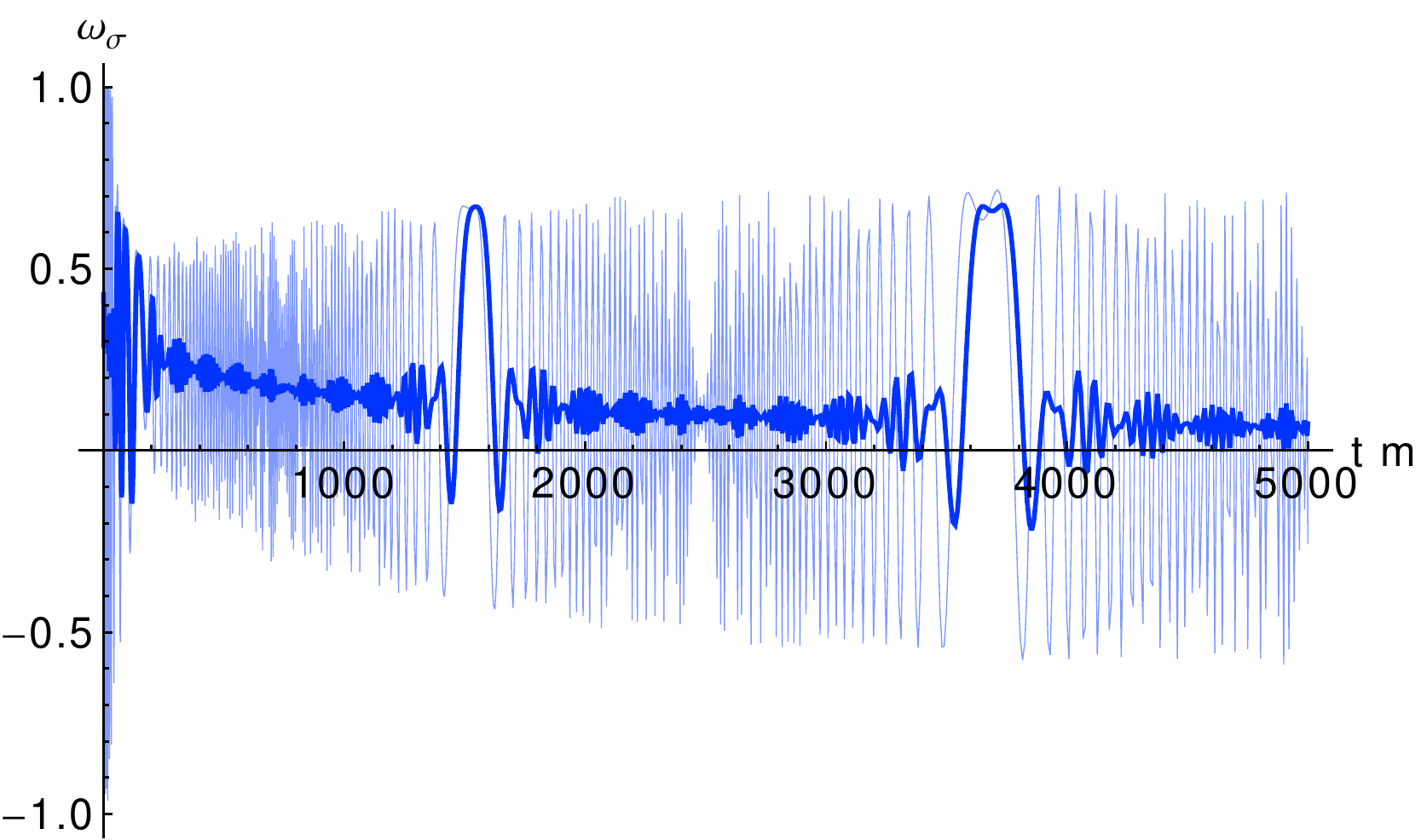}}
\caption{The evolution of the equation of the state of the curvaton $\omega_{\sigma}$ \ie Eq. (\ref{eqs}) during the simulation. Note that the thin line is the variable and the thick line is the conformal moving average. Notice also that the large gaps at $t_\textrm{phys}/m \sim 1400$ and $t_\textrm{phys}/m \sim 3800$ in the graph are an artifact of the output sampling.}
\label{fig5}
\end{figure}

\begin{figure}[h]
{\includegraphics*[width=\columnwidth]{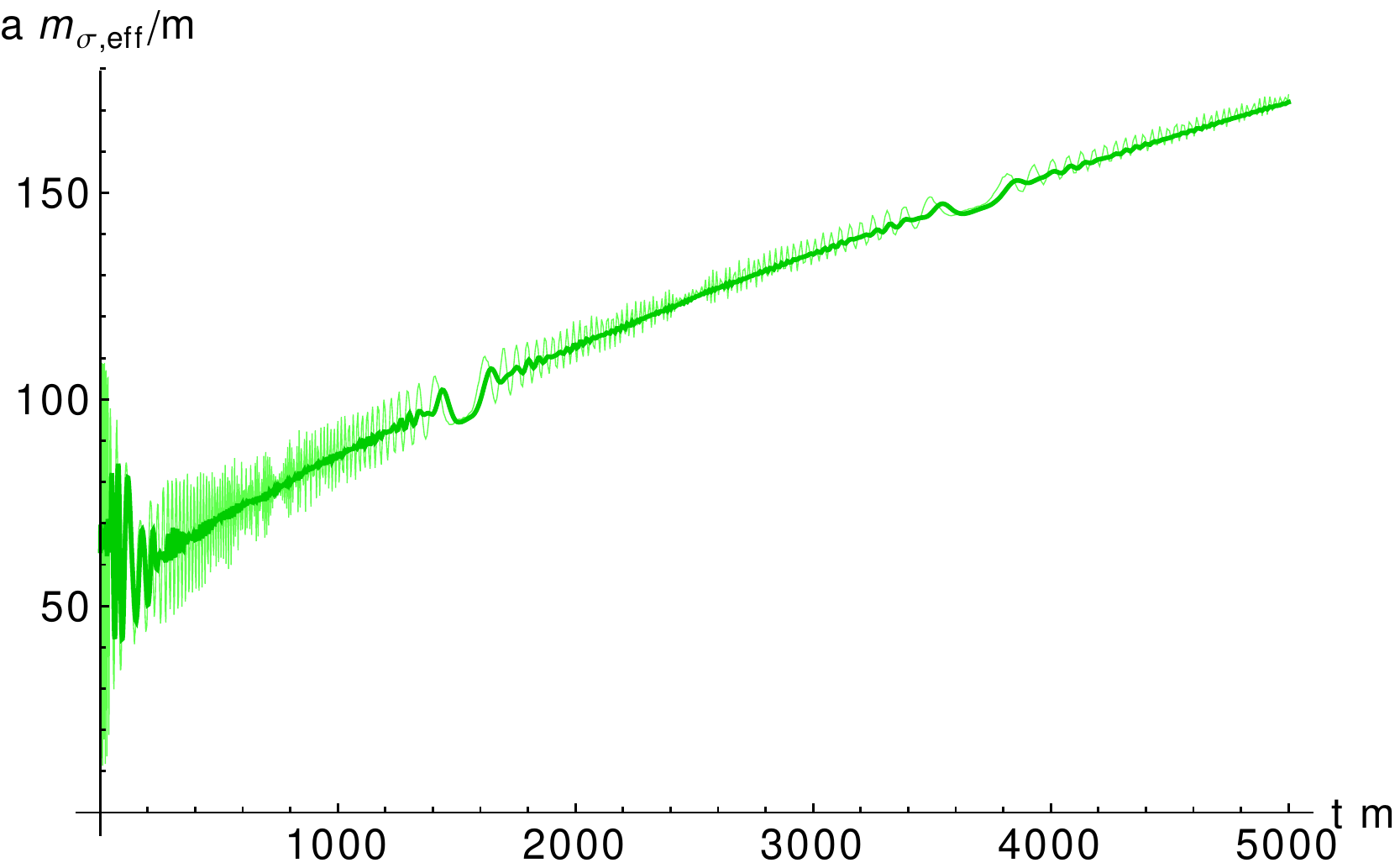}}
\caption{The evolution of the comoving effective mass $a \; m_{\textrm{eff}}$ during the simulation. Note that the thin line is the variable and the thick line is the conformal moving average.}
\label{fig6}
\end{figure}

\begin{figure}[h]
{\includegraphics*[width=\columnwidth]{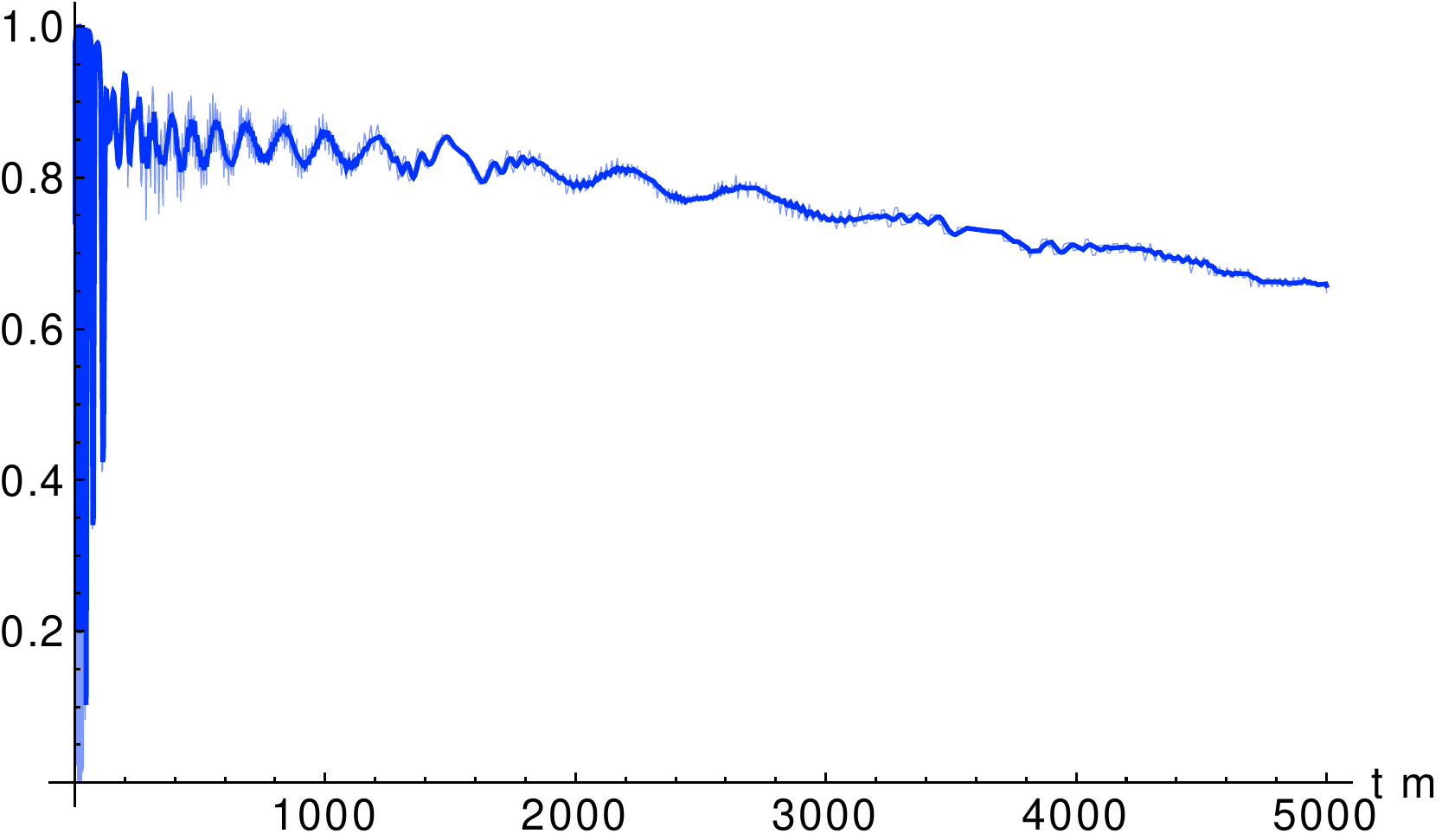}}
\caption{The evolution of the fraction of relativistic curvaton particles, \ie for which $k > a \; m_\textrm{eff}$, during the simulation. Note that the thin line is the variable and the thick line is the conformal moving average.}
\label{fig7}
\end{figure}

We are also interested in the evolution of the equation of the state of the curvaton during the thermalization process. As can be seen from Figure \ref{fig5} the system is initially highly relativistic and oscillates rapidly. This oscillatory phase corresponds to the exponential increase in the comoving particle number density seen in Figure \ref{fig2}. As the system evolves the equation of state starts to decrease but at the end of simulation its average is still close to a value of $0.1$ indicating that the system is not yet non-relativistic.

In Figure \ref{fig6} we plot the comoving effective mass $a \; m_\textrm{eff}$ in units of $m$ which we calculate as an average over the lattice. The early stages are in this case also highly oscillatory which is followed by a period of gradual increase due to the expansion of the universe. 
During the resonance and the rescattering periods the comoving effective mass stays almost constant and it starts to grow only after the mass term starts to dominate at $t_\textrm{phys} \, m \sim 200$.

In Figure \ref{fig7} we show the fraction of curvaton particles that are relativistic \ie for which $k > a \; m_\textrm{eff}$ and the homogeneous mode is not included in the calculations. The figure shows that the created curvaton particles are highly relativistic during the simulation with a final value close $65$ percent. Notice that the discrepancy between Figures \ref{fig5} and \ref{fig7} is caused by the coherent curvaton field that still gives a significant contribution to the energy and pressure densities of the curvaton component at the end of the simulation.

\begin{figure}[h]
{\includegraphics*[width=\columnwidth]{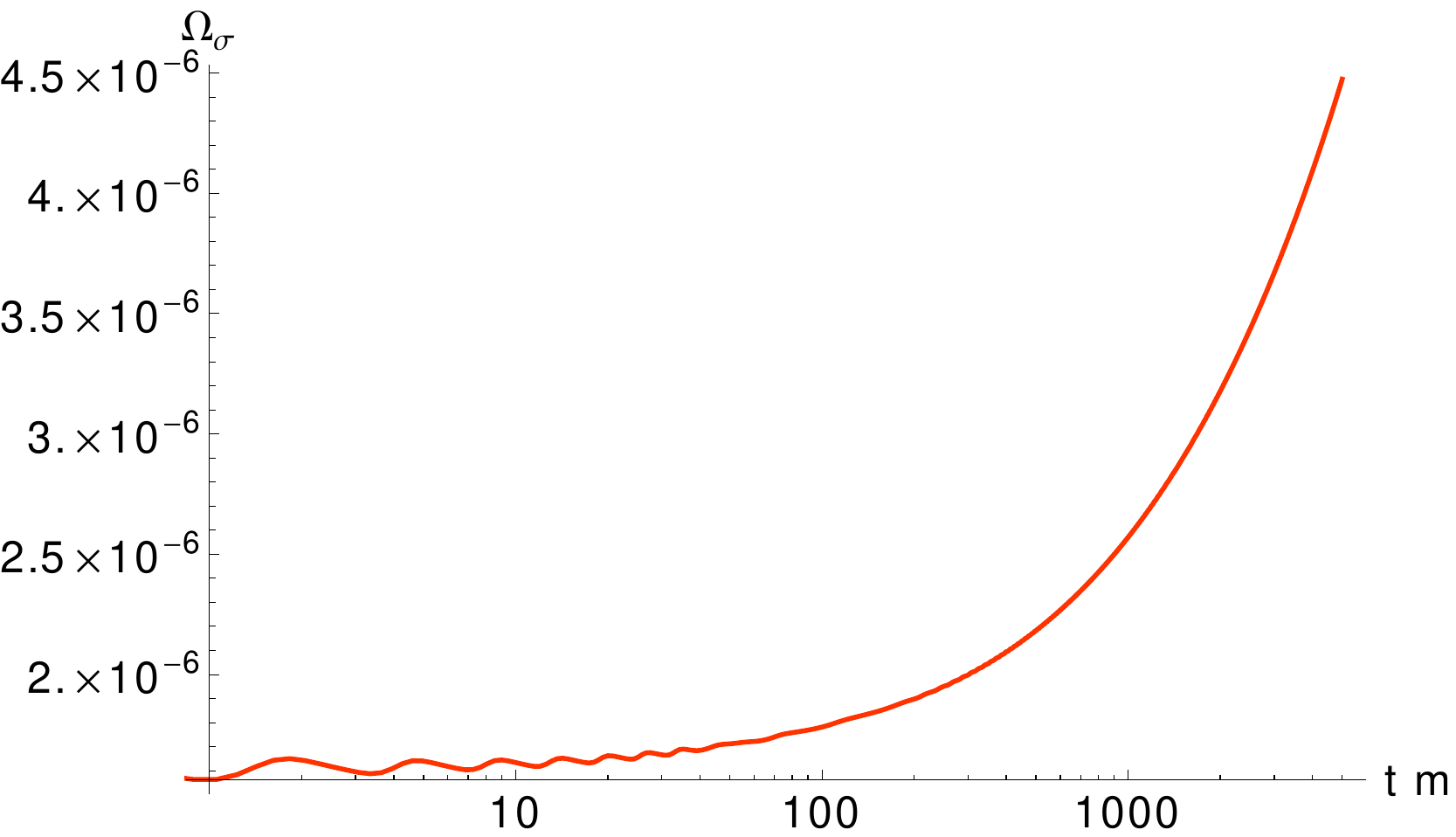}}
\caption{The evolution of the fractional energy density of the curvaton $\Omega_{\sigma}$ as a function of time.}
\label{fig8}
\end{figure}

Yet another aspect of the evolution of the curvaton is seen in Figure \ref{fig8} where we plot the fractional energy density of the curvaton during the simulation. Initially it evolves in tandem with the homogeneous radiation component up to time $t m \sim 10$ after which its fraction of energy density starts to grow steadily as its equation of state starts to approach that of matter.

\begin{figure}[h]
{\includegraphics*[width=\columnwidth]{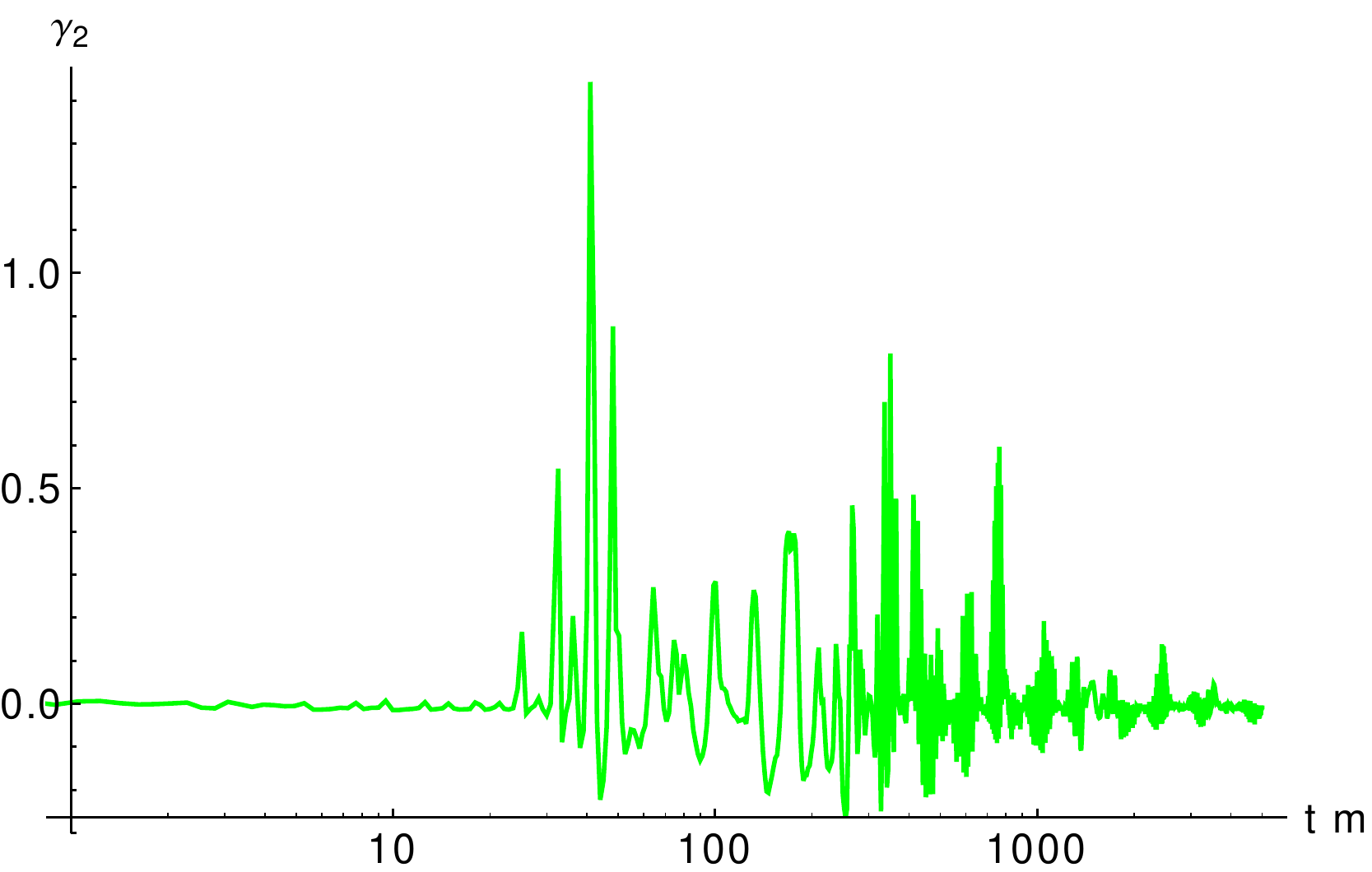}}
\caption{The evolution of excess kurtosis \ie Eq. (\ref{kurtosis}) during the simulation. Note that time is given in logarithmic units.}
\label{fig9}
\end{figure}

We have also included a plot of the excess kurtosis $\gamma_2$ during the thermalization process in Figure \ref{fig9}. In the early highly oscillatory preheating phase the system is also highly non-gaussian. However as the simulation progresses the curvaton field starts to return to gaussian. This behavior is very similar to the one observed in the parametric resonance of the chaotic inflaton field \cite{Felder:2006cc}. The skewness of the curvaton field shows a very similar trend and we have therefore omitted the graph of its evolution.

\subsection{Non-gaussianity calculations}

The possible generation of non-gaussianity during the curvaton thermalization process is an interesting and a timely question in cosmology \cite{Komatsu:2003fd,Bartolo:2004if,Komatsu:2008hk,Byrnes:2010em}. To calculate this we will use the $\Delta N$ formalism based on the separate universe approach \cite{Wands:2000dp} that has been previously applied successfully to different parametric resonance scenarios \cite{Chambers:2007se,Chambers:2008gu,Bond:2009xx,Chambers:2009ki}. In the separate universe approach different patches of the universe that are separated by more than a Hubble distance are presumed to evolve independently of each other. Assuming also that each Hubble volume is isotropic and homogeneous they can be approximated to be separate Friedmann-Robertson-Walker 'universes'. The evolution of these patches is solved with the lattice simulation method as in the previous section.

The curvature perturbation on scales larger than the Hubble horizon is defined as
\begin{equation} \label{zeta}
\zeta = \delta \ln a |_{H},
\end{equation}
where the difference in the scale factor is calculated at a hypersurface of constant Hubble parameter $H$. The scale factor is normalized to be one at the start of the curvaton thermalization process. We will vary the homogeneous value of the curvaton field with superhorizon fluctuations from one patch to another. This will cause slight variations in the value of the curvature perturbation $\zeta$. For small perturbations $\delta \sigma$ equation (\ref{zeta}) is often expanded as
\begin{equation} \label{zeta-2}
\zeta = (\ln a)' \Big|_{H} \delta \sigma + \frac{1}{2}\ln a'' \Big|_{H} \delta \sigma^2 + \dots ,
\end{equation}
where the primes are derivatives calculated with respect to the curvaton value at the end of inflation on hypersurfaces of constant Hubble parameter $H$.
The spectrum of the curvature perturbation can be written with this as
\begin{equation} \label{zeta spectra}
P_\zeta = \Big[\ln a'\Big]^2 P_\sigma,
\end{equation}
where $P_\sigma$ is the spectrum of the curvaton field. Following \cite{Chambers:2008gu,Chambers:2009ki} we will use
\begin{equation} \label{sigma spectrum}
P_\sigma(k) \approx \frac{H_k^2}{4 \pi^2} \approx \frac{4}{3 \pi^2}\lambda m^2_{\textrm{PL}} N_k^2,
\end{equation}
which is valid for massless fields during inflation. $N_k (\approx 60)$ here measures how many number of e-foldings before the end of inflation mode $k = a H_k$ left the Hubble horizon. The local non-gaussinity parameter can be defined also in terms of the coefficients of equation (\ref{zeta-2}) \cite{Lyth:2005fi} as
\begin{equation} \label{f NL}
f_{\textrm{NL}} = \frac{5}{6}\frac{\ln a''}{(\ln a)'^2}\Bigg|_{H}.
\end{equation}

To calculate the non-gaussianity in the curvaton scenario we will apply the method presented in \cite{Chambers:2009ki} with minor modifications. We will write the energy density as a combination of the relativistic radiation and the curvaton component which we assume to behave like matter 
\begin{equation} \label{rho}
\rho = \rho_{\textrm{ref}}\Big[r_{\textrm{ref}}\Big(\frac{a}{a_{\textrm{ref}}}\Big)^{3}+(1-r_{\textrm{ref}})\Big(\frac{a}{a_{\textrm{ref}}}\Big)^{4}\Big],
\end{equation}
where the fractional energy density of curvaton $r_{\textrm{ref}} = \Omega_{\sigma,\textrm{ref}}$, scale factor $a_{\textrm{ref}}$ and energy density $\rho_{\textrm{ref}}$ are calculated at a reference point defined after the resonance period of the curvaton. 

We will assume that the curvaton stays subdominant during its evolution and decays perturbatively when the Hubble parameter $H$ is of the order of the decay width $\Gamma$. We will use the sudden decay approximation by assuming that this decay is instantaneous. The value of the decay width is unknown meaning that the energy density $\rho_{\textrm{decay}}$ and the fractional energy density $r_{\textrm{decay}}$ at the moment of decay are free parameters limited by observational data, namely the amplitude of the curvature perturbations.
By now taking logarithms on both side of equation (\ref{rho}), expanding the right side in series
with respect to $r_{\textrm{ref}}$ and rearranging the terms the logarithm of the scale factor reads
\begin{equation} \label{ln_a}
\begin{aligned}
\ln a = & \ln a_{\textrm{ref}} + \frac{1}{4}\Bigg[ \ln \frac{\rho_{\textrm{ref}}}{\rho} + r - r_{\textrm{ref}}\Bigg] \\
= & \ln a_{\textrm{ref}} + \frac{1}{4}\Bigg[ \ln \frac{\rho_{\textrm{ref}}}{\rho} + C r_{\textrm{ref}}\Bigg],
\end{aligned}
\end{equation}
where
\begin{equation}
r \equiv r_{\textrm{ref}}\Big(\frac{\rho_{\textrm{ref}}}{\rho}\Big)^{1/4}
\end{equation}
and $C \equiv r/r_{\textrm{ref}} -1$. The curvature perturbation can now be written as
\begin{equation} \label{zeta-eq}
\zeta(\hat{\sigma}_0) = \ln a(\hat{\sigma}_0) - \ln a(\sigma_0) = \delta \ln a_{\textrm{ref}} + C \delta r_{\textrm{ref}}
\end{equation}
where we have written explicitly the dependence on the curvaton value at the end of inflation.  We have also neglected the energy density terms from equation (\ref{ln_a}) since the calculations are done on a constant $H$ hypersurface on which also the energy density is constant by the Friedmann equations.

We will now assume that the logarithm of the scale factor and the fractional energy density can be expanded in terms of the superhorizon fluctuations of the homogeneous curvaton values similarly to equation (\ref{zeta-2}):
\begin{equation} \label{ln_a_r}
\begin{aligned}
\ln a_{\textrm{ref}}(\hat{\sigma}_0) = & \ln a_{\textrm{ref}}(\sigma_0) + \ln a_{\textrm{ref}}'(\hat{\sigma}_0 - \sigma_0) \\
&  + \frac{1}{2}\ln a_{\textrm{ref}}''(\hat{\sigma}_0 - \sigma_0)^2,\\
r_{\textrm{ref}}(\hat{\sigma}_0) = & r_{\textrm{ref}}(\sigma_0) + r_{\textrm{ref}}'(\hat{\sigma}_0 - \sigma_0) \\
& + \frac{1}{2}r_{\textrm{ref}}''(\hat{\sigma}_0 - \sigma_0)^2,\\
\end{aligned}
\end{equation}
where $\hat{\sigma}_0 = \sigma_0 + \delta\sigma_0$ and $\delta\sigma_0$ is a superhorizon fluctuation of the initial curvaton value. Equations (\ref{ln_a_r}) are fitted to the simulation data to get numerical values for the polynomial coefficients $\ln a_{\textrm{ref}}'$, $\ln a_{\textrm{ref}}''$, $r_{\textrm{ref}}'$ and $r_{\textrm{ref}}''$.
For the amplitude of the curvature perturbation spectrum (\ref{zeta spectra}) to be consistent with the WMAP observations \cite{Spergel:2006hy}, $P_{\zeta} \simeq 2.4\times 10^{-9}$, the unknown fractional energy density of the the curvaton at the moment of decay can be solved \cite{Chambers:2009ki} in terms of the power spectrum amplitudes and the polynomial coefficients
\begin{equation} \label{r-decay}
r_{\textrm{decay}} = r_{\textrm{ref}} + 4\frac{r_{\textrm{ref}}}{r_{\textrm{ref}}'}\Bigg(\pm \sqrt{\frac{P_\zeta}{P_\sigma}} - \ln a_{\textrm{ref}}' \Bigg).
\end{equation}
The non-gaussianity parameter (\ref{f NL}) can be written similarly \cite{Chambers:2009ki} as
\begin{equation} \label{f NL2}
f_{\textrm{NL}} = \frac{5}{6}\frac{P_\sigma}{P_\zeta}\Bigg(\ln a_{\textrm{ref}}'' + \frac{r_{\textrm{ref}}''}{r_{\textrm{ref}}'}\Bigg(\pm \sqrt{\frac{P_\zeta}{P_\sigma}} - \ln a_{\textrm{ref}}' \Bigg)\Bigg).
\end{equation}

\begin{figure}[h]
{\includegraphics*[width=\columnwidth]{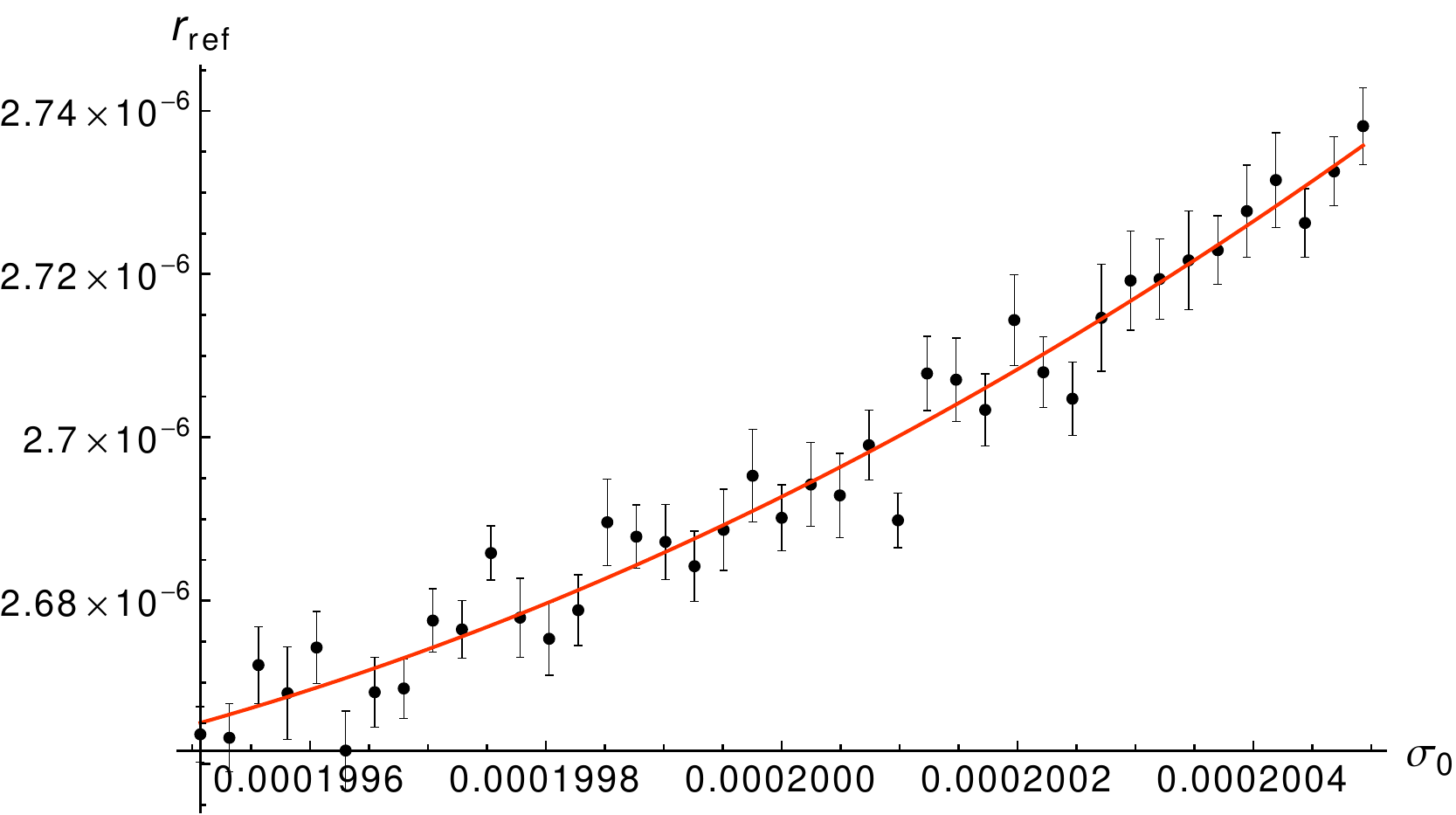}}
\caption{The fractional energy density $r_{\textrm{ref}}$ calculated at the reference value of the Hubble parameter $H_{\textrm{ref}}$ as a function of the initial homogeneous value of the curvaton. The continuous line in the graph represents a second order polynomial least squares fit to the data. We have also included standard error limits of the simulation data.}
\label{fig10}
\end{figure}

\begin{figure}[h]
{\includegraphics*[width=\columnwidth]{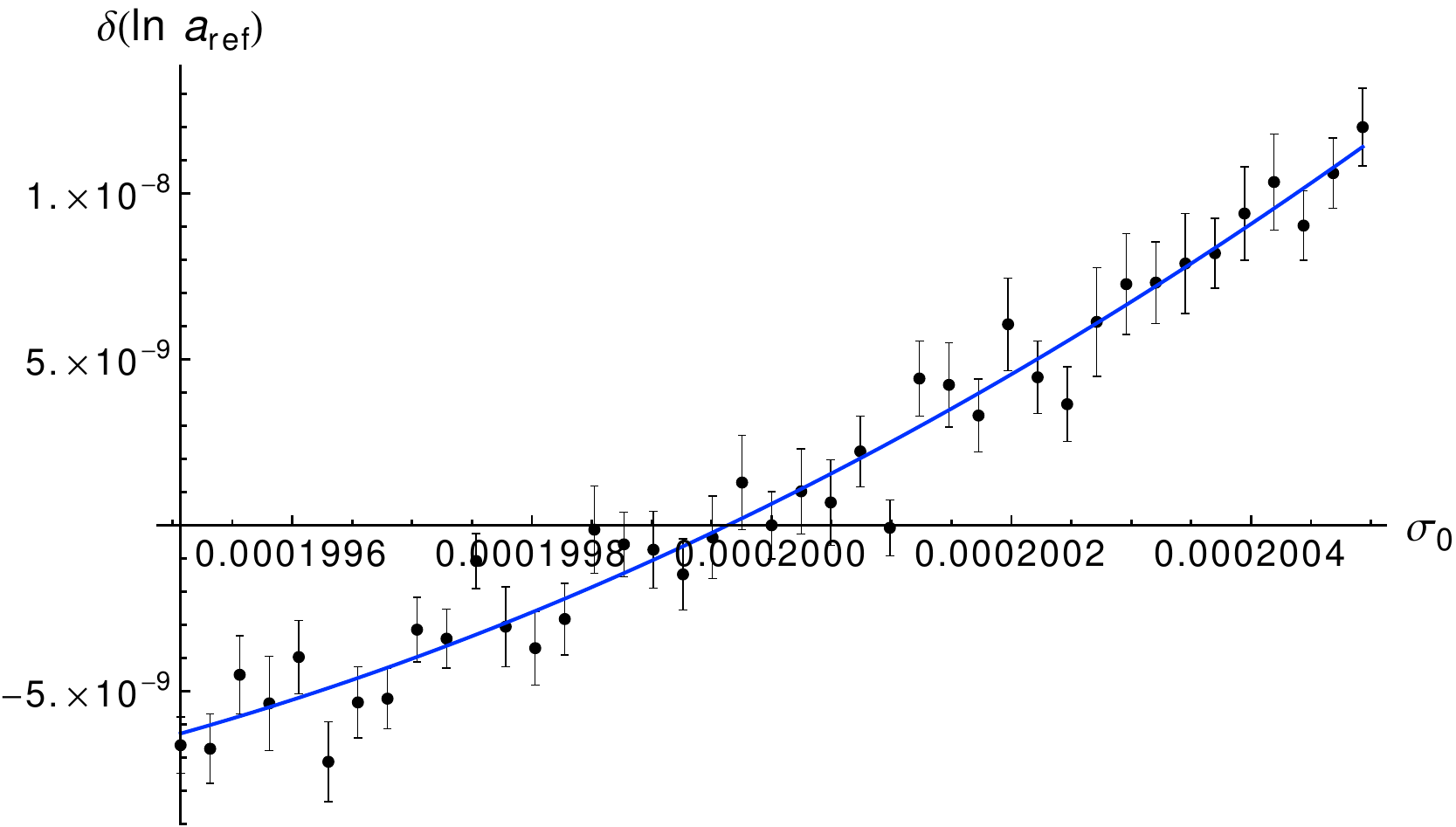}}
\caption{The difference of the logarithm of the scale factor $a$ calculated at 
$H_{\textrm{ref}}$ as a function of the initial homogeneous value of the curvaton. The continuous line in the graph represents a second order polynomial least squares fit to the data. We have also included standard error limits of the simulation data.}
\label{fig11}
\end{figure}

The Monte Carlo simulations were run with the initial values that were used in the thermalization analysis. As mentioned previously we used a smaller lattice size of $64^3$ points in order to shorten the overall simulation runtime drastically (roughly 46 times faster). For the reference point where the different quantities are calculated we use $H_{\textrm{ref}} = 4 \times 10^{-4} m$ which in terms of physical time corresponds to $t m \simeq 1260$. The actual value is determined by interpolating around $H_{\textrm{ref}}$.
The range of homogeneous curvaton values over which the simulations need to be run is determined by the variance of the curvaton values at the end of inflation. For inflation potential (\ref{V phi}) and curvaton spectrum (\ref{sigma spectrum}) this reads \cite{Chambers:2009ki}
\begin{equation}
\langle \delta \sigma^2 \rangle \approx \frac{4}{9\pi^2}\lambda m^2_{\textrm{PL}} N_0^3
\end{equation}
where $N_0 \approx 60$ is the number of e-foldings after the largest currently observable scales left the horizon.
The range of curvaton initial values then reads
\begin{equation}
\sigma_0 - \frac{1}{2}\delta \sigma_0 \leq \hat{\sigma}_0 \leq \sigma_0 + \frac{1}{2}\delta \sigma_0,
\end{equation}
where $\delta \sigma_0 = \sqrt{\langle \delta \sigma^2 \rangle} \approx 9.9 \times 10^{-7}$ and $\sigma_0 = 0.0002$. We take 41 equidistant points from this range and use as the initial homogeneous curvaton values. At each point the simulations are solved with different random field perturbations 35 times to get the necessary statistics. Note that these subhorizon perturbations are generated with a convolution based algorithm presented in \cite{Frolov:2008hy}. The total simulation runtime with these selections is roughly 25 hours when using a Nvidia Tesla C2050 computing card.

\begin{figure}[h]
{\includegraphics*[width=\columnwidth]{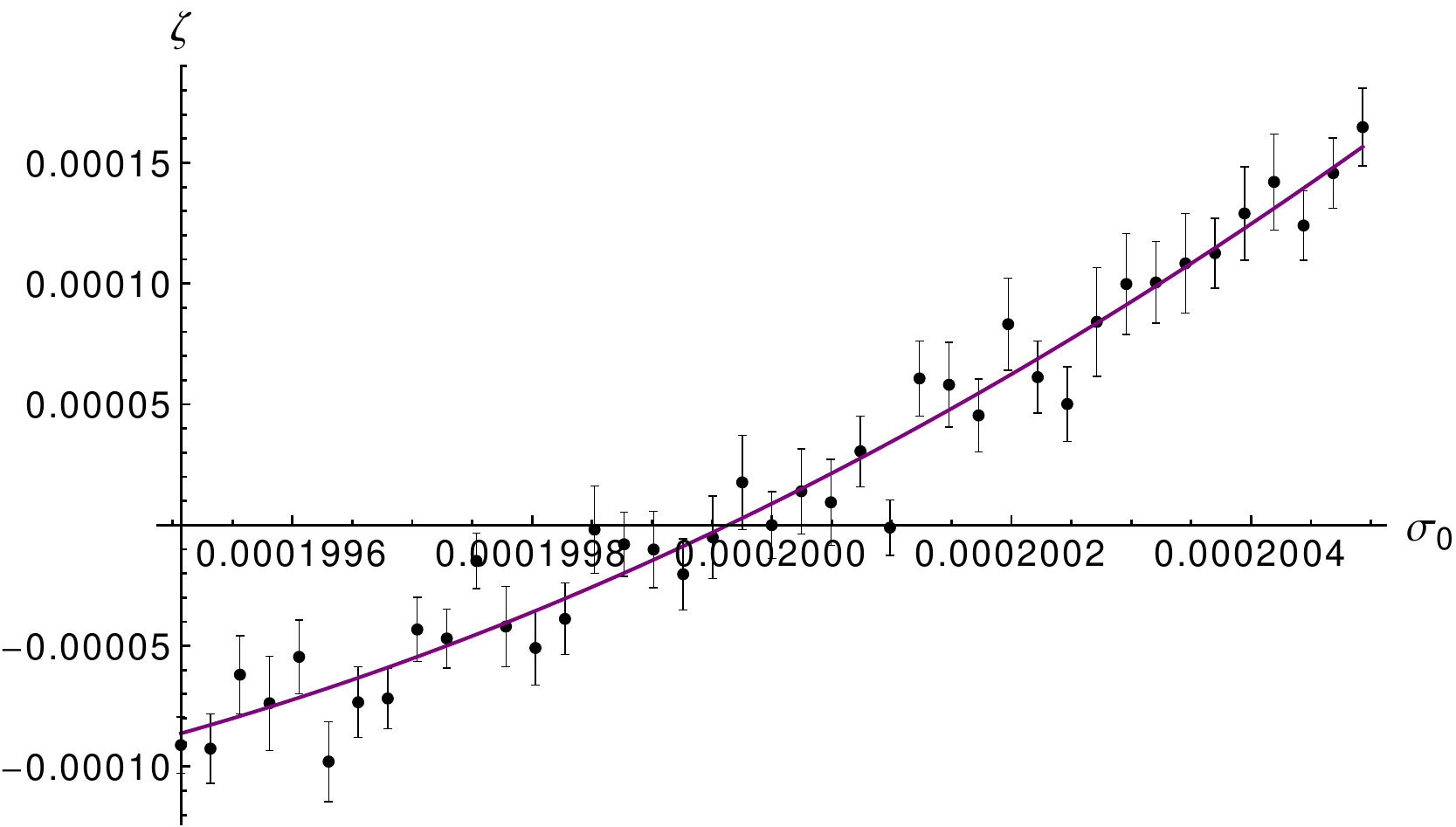}}
\caption{The curvature perturbation $\zeta$ as a function of the initial homogeneous value of the curvaton. The continuous line in the graph represents a second order polynomial least squares fit to the data. We have also included standard error limits of the simulation data.}
\label{fig12}
\end{figure}

\begin{center}
\begin{table*}[ht]
\begin{tabular}{|c|c|c|c|}
\hline
 & 1 & $\hat{\sigma}_0 - \sigma_0$ & $(\hat{\sigma}_0 - \sigma_0)^2$ \\
\hline
\hline
$\ln a_{\textrm{ref}}(\hat{\sigma}_0) - \ln a_{\textrm{ref}}(\sigma_0)$ & $(6.465 \pm 5.058) \times 10^{-10} $ & $(1.793 \pm 0.1155 )\times 10^{-2}$ & $7885  \pm 4430$ \\
\hline
$r_{\textrm{ref}}(\hat{\sigma}_0)$ & $(2.693 \pm 0.002022)\times 10^{-6}$ & $(7.170 \pm 0.4620 )\times 10^{-2}$ & $ 31530 \pm 17720 $\\
\hline
$\zeta$ & $(8.868 \pm 6.947)\times 10^{-6}$ & $246.3  \pm 15.87$ & $(1.083 \pm 0.6086)\times 10^{8}$\\
\hline
\end{tabular}
\caption{List of least squares fit results for second order polynomials $\ln a_{\textrm{ref}}(\hat{\sigma}_0)$, $r_{\textrm{ref}}(\hat{\sigma}_0)$ and $\zeta$ \ie equations (\ref{ln_a_r}) and (\ref{zeta-2}) respectively. In the columns we have given the coefficients of different powers of $\hat{\sigma}_0 - \sigma_0$. We have also given the confidence intervals for the parameters at $95$ \% level.}
\label{table-1}
\end{table*}
\end{center}

The main results of the simulations are presented in Figures \ref{fig10}-\ref{fig12} and in Table \ref{table-1}. In Figures \ref{fig10} and \ref{fig11} we have the fractional energy density $r_{\textrm{ref}}$ and the difference of the logarithm of the scale factor $a$ calculated at the reference value of the Hubble parameter $H_{\textrm{ref}}$ as a function of the initial homogeneous value of the curvaton. The curvature perturbation $\zeta$ calculated with formula (\ref{zeta-eq}) is given in Figure \ref{fig12}. We have also included least square fits of the equations (\ref{ln_a_r}) in the graphs with the corresponding polynomial coefficients given in Table \ref{table-1}. Note that we have also included the confidence intervals of the parameters at 95 \% level which were derived from the fitting results given by Mathematica.

With these results the curvaton fraction at decay reads $r_{\textrm{decay}} = 0.037 \pm 0.0024$. Assuming that the radiation stays dominant after the end of the simulation and that $\Omega_{\sigma} \simeq \rho_{\sigma}/\rho_{\gamma} \sim a$ during this period the value of the Hubble parameter at decay can be calculated to be roughly $0.1 \textrm{ eV}$ which translates to a reheating temperature $T_{\textrm{reh}} \sim 1 \textrm{ GeV}$ which is considerably higher than the result of the previous section. The non-gaussianity variable can be calculated from equation (\ref{f NL2}) or by fitting $\zeta$ directly with equation (\ref{zeta-2}). The results are $f_{\textrm{NL}} = 2980 \pm 644$ and $f_{\textrm{NL}} = 2976 \pm 1481$ respectively at $95 \%$ confidence level. When compared to the results of a two field curvaton resonance model \cite{Chambers:2009ki} the quadratic polynomials follow more closely the general trend of the data. Despite this the calculated level of non-gaussianity is still very high and the current observational limit \cite{Komatsu:2008hk} $-9 < f_{\textrm{NL}} = 111$ at $95 \%$ confidence level rules out  the model with the current parameter values.

This large level of non-gaussianity is mainly caused by the magnitude of the second order coefficient $r_{\textrm{ref}}''$ and the smallness of the first order coefficient $r_{\textrm{ref}}'$ in equation (\ref{f NL2}). An easy remedy to this would be to use a smaller curvaton self-interaction strength which would lead to a more linear evolution of the fractional energy density of the curvaton in Figure \ref{fig10}. This might however cause some thermalization related problems mentioned briefly in the previous section: for smaller values of self-interaction coupling strength the rescattering phase after the resonance period was found to be very weak and limited and the final shape of the number spectrum exhibit a clear peak at $k \sim k_p$. The created comoving number density of the particles in this case would be also orders of magnitude smaller than with the current values.

\section{Discussion and conclusions}

We have studied the self-interacting curvaton scenario with classical fields and lattice simulations from two different perspectives. First we concentrated on the thermalization process during and after the preheating phase. The results indicate that in the current curvaton scenario the overall evolution of system follows closely the previously studied self-interacting inflaton model. We found that during the resonance period curvaton particles were created at a predicted resonance band and in the ensuing rescattering phase the spectrum developed peaks at harmonic frequencies related to the momentum values of the resonance band.
The final state of the curvaton field could be characterized as a pre-thermalized one.

After this we concentrated on the calculation of the generated non-gaussianity during the resonance. We employed and adapted a previously presented method \cite{Chambers:2009ki} to the self-interacting curvaton scenario. When compared to the broad resonance of curvaton \cite{Chambers:2009ki} the simulation data was found to be a better fit to the used quadratic approximation of the curvature perturbation. The used parameter values were however rule out by the current observational limits and were found to be unphysical. There might however be regions in the parameter space that could lead to non-gaussianities consistent with the observations. This would however take more computing resources that were available while doing this paper.
One option would be to make a distributed version of PyCOOL that would systematically scan the parameter space for suitable initial values.
Another interesting possibility would be to study the generation of gravitational waves during the curvaton resonance which would likely give additional limitations on the curvaton model. This could be done easily with a recently updated version of PyCOOL and we leave it for future work.

\subsection*{Acknowledgments}

The author is thankful to Arttu Rajantie for useful comments and for providing a code that was helpful when developing the non-gaussianity procedures.
Useful comments and discussions with Iiro Vilja are also gratefully acknowledged.



\end{document}